\documentclass[aps,prl,prabib,amsmath,nofootinbib,
amssymb,amsfonts,smallcaptions,showpacs]{revtex4}
\usepackage{psfig}
\usepackage{graphicx}

\newcommand{\eqname}[1]{\label{eq:#1}}
\newcommand{\bgar}{\begin{eqnarray}}
\newcommand{\enar}[1]{\label{eq:#1}\end{eqnarray}}
\newcommand{\valass}[1]{\left|#1\right|}
\newcommand{\norme}[1]{\left\|#1\right\|}
\newcommand{\trace}[1]{\textrm{Tr}\left[#1\right]}
\newcommand{\ket}[1]{ | #1 \rangle }

\newcommand{\braket}[2]{ \langle #1 | #2 \rangle }

\newcommand{\ketbra}[2]{ | #1 \left\rangle \right\langle #2 |}

\newcommand{\mean}[1]{\overline{#1}}
\newcommand{\expect}[1]{\left\langle #1 \right\rangle}

\newcommand{\kk}{ {\bf k}}
\newcommand{\rr}{ {\bf r}}

\newcommand{\proj}{{\mathcal Q}}

\newcommand{\eq}[1]{(\ref{eq:#1})}

\newcommand{\Psihd}{\hat\Psi^\dagger}

\newcommand{\Psih}{\hat\Psi}

\newcommand{\Hamilt}{{\mathcal H}}
\newcommand{\ahd}{\hat a^\dagger}
\newcommand{\ah}{\hat a}

\begin{document}

\title{Exact reformulation of the bosonic many-body
problem in terms of stochastic wave functions: convergence issues}

\affiliation{Laboratoire Kastler Brossel, \'Ecole Normale
Sup\'erieure, 24 rue Lhomond, 75231 Paris Cedex 05, France}

\author{Iacopo Carusotto}
\affiliation{Laboratoire Kastler Brossel, \'Ecole Normale
Sup\'erieure, 24 rue Lhomond, 75231 Paris Cedex 05, France}

\author{Yvan Castin}
\email{Yvan.Castin@lkb.ens.fr}
\affiliation{Laboratoire Kastler Brossel, \'Ecole Normale
Sup\'erieure, 24 rue Lhomond, 75231 Paris Cedex 05, France}

\begin{abstract}
There exist methods to reformulate in an exact way
the many-body problem of interacting bosons in terms 
of the stochastic evolution of single particle wave functions.
For one such reformulation, the so-called {\em simple} Fock
scheme, we present an elementary derivation, much simpler than the
original one.
Furthermore, we show that two other schemes, based on coherent
states of the matter field rather than on Fock states, lead
to an infinite statistical uncertainty in
the continuous time limit.
The {\em simple} Fock scheme is therefore, up to now, the only
one that was proved to lead to a convergent Monte Carlo simulation
scheme at all times.
\end{abstract}


\pacs{05.30.Jp, 03.75.Fi, 02.70.Ss  }


\date{\today}

\maketitle



\section{Introduction}
Consider a gas of $N$ indistinguishable bosonic spinless particles
with binary interactions, a situation typically encountered
in Bose condensed atomic gases \cite{RevueBEC}.
The time evolution of the many-body state vector $\ket{\psi}$ of
the gas is described by the Schr\"odinger equation
\begin{equation}
\frac{d}{dt}\ket{\psi}=\frac{1}{i\hbar}\Hamilt\ket{\psi}
\eqname{Schrod}
\end{equation}
where $\Hamilt$ is the Hamiltonian of the system, including
kinetic energy, trapping potential energy and interaction terms.
In the general case of a large number of particles
and a large number of modes of the atomic matter field,
the Hilbert space of the system is far too large 
for a direct integration of \eq{Schrod} to be feasible on a computer.
A possible way out, intensively explored in particular in the quantum
optics community, is to reformulate the evolution equation in terms 
of the stochastic evolution of classical fields, which are much
smaller objects. Historically, the first suggestion for such a
reformulation is based on the Positive-P distribution
\cite{PosP1,PositiveP}. Other schemes have been proposed more
recently, based on different representations of the many-body
density operator \cite{GPstoch,Plimak,Deuar}. Generalizations
to fermionic systems are also possible \cite{Plimak_fermion,Chomaz}.

As these methods are stochastic in nature, their prediction for a
finite number of realizations deviates from the exact result by a
fluctuating, zero-mean quantity, called the {\em statistical error}.
One of the challenges of such reformulations is to prove that the
average of the absolute value of the statistical error remains finite
at all times, so that an actual Monte Carlo simulation converges to
the exact result in the limit of an infinite number of realizations
\footnote{Let us call $\delta$ the absolute value of the
statistical error (to be defined more precisely later on). 
Let us assume that the probability distribution for $\delta$ has a
long tail scaling as $1/\delta^{1+\mu}$ for large
$\delta$. The most favorable case is $\mu>2$, in which case $\delta$
has a finite variance and the statistical error for a sample of
${\mathcal N}_r$ Monte Carlo realizations scales as $1/\sqrt{{\mathcal
N}_r}$. In the intermediate $1<\mu<2$ case, the convergence is slower,
scaling as $1/{\mathcal N}_r^{1-1/\mu}$. In the worst $\mu<1$ case,
the mean value of $\delta$ is infinite and the statistical error of
the Monte Carlo simulation grows to infinity with the number of
realizations as ${\mathcal N}_r^{-1+1/\mu}$. For more details, 
see~\cite{Levy}.}.
To our knowledge, a proof of such a convergence property for the
multi-mode Hamiltonian problem defined above was given only 
for the so-called {\em simple} Fock scheme in \cite{GPstoch}
and for its fermionic counterpart \cite{Chomaz}.

The goal of the present paper is twofold. First, it presents
a much simpler derivation of the {\em simple} Fock scheme 
than the original one in \cite{GPstoch}: the main idea here
is to work directly with state vectors $|\psi\rangle$ rather than with
the many-body density operator. Second, it addresses the issue
of the convergence of two other schemes based on coherent states
of the bosonic field: the {\em simple} coherent scheme of \cite{GPstoch}
and a new scheme that we call the Bargmann scheme. It is found that
these two schemes based on coherent states lead to an infinite mean of
the absolute value of the statistical error for any finite evolution
time.

\section{The model Hamiltonian}

A model for an ultracold trapped interacting Bose gas in $D$
dimensions with short range interactions can be obtained in a second-quantization formalism by using
the Hamiltonian
\begin{equation}
\label{eq:Hamilt}
{\mathcal H}=\sum_\rr \Delta V\,\Psihd(\rr)h_{\rm 0}\Psih(\rr)+
\frac{g_0}{2} \sum_\rr \Delta V\,\Psihd(\rr)\Psihd(\rr)\Psih(\rr)\Psih(\rr);
\end{equation}
the spatial coordinate $\rr$ runs on a discrete orthogonal lattice 
of $M$ points with periodic boundary conditions; $V$ is the
total volume of the quantization box and $\Delta V=V/M$ is the
volume of the unit cell of the lattice.
$h_0=\frac{p^2}{2m}+U_{\rm ext}$ is the one-body Hamiltonian in the
external trapping potential
$U_{\rm ext}(\rr)$, $m$ is the atomic mass and interactions are
modeled by a two-body discrete delta potential with a coupling
constant $g_0$.
The field operators $\Psih(\rr)$ satisfy the Bose commutation
relations $[\Psih(\rr),\Psihd(\rr')]=\delta_{\rr,\rr'}/\Delta V$ and
can be expanded on plane waves according to $\Psih(\rr)=\sum_\kk
\ah_\kk e^{i\kk\rr}/\sqrt{V}$ with $\kk$ restricted to the
first Brillouin zone of the reciprocal lattice; in the plane wave
basis, the kinetic energy term has the diagonal form
$\sum_\kk\frac{\hbar^2 \kk^2}{2m}\ahd_\kk \ah_\kk$.
In order for the discrete model
to correctly reproduce the underlying continuous field theory, the
grid spacing must be smaller than macroscopic length scales
like the thermal de Broglie wavelength and the healing length.

\section{The {\em simple} Fock scheme}
\label{sec:SF}

Consider the $N$ particle bosonic Hartree-Fock state defined as
\begin{equation}
\ket{N:\phi}=\frac{1}{\sqrt{N!}}
\ah_\phi^{\dagger N}\ket{0}
\eqname{Hartree}
\end{equation}
where the operator $\ah_\phi^{\dagger}$ creates a particle in the
non-necessarily normalized single-particle wave function $\phi(\rr)$:
\begin{equation}
\ah_\phi^{\dagger}=\sum_{\rr} \Delta V\,\phi(\rr)\Psihd(\rr).
\end{equation}
The challenge is here to find an equation of motion for $\phi(\rr)$ in
order for the Hartree-Fock ansatz to give an exact solution to the
full many-body dynamics, i.e. we aim to obtain a dynamics for $\phi(\rr)$ 
such that
\begin{equation}
d\ket{N:\phi}\stackrel{?}{=}\frac{dt}{i\hbar}\Hamilt\ket{N:\phi}
\eqname{???}
\end{equation}
in a sense to be defined.
In the following we proceed to evaluate both sides of \eq{???}.

The action of the Hamiltonian \eq{Hamilt} on $\ket{N:\phi}$ can be
worked out by splitting the field operator in its longitudinal and
orthogonal components with respect to $\phi$:
\begin{equation}
\Psih(\rr)=\frac{\phi(\rr)}{\norme{\phi}^2}\,\ah_\phi+\Psih_\perp(\rr)
\end{equation}
where the squared norm of $\phi$ is 
\begin{equation}
\norme{\phi}^2=\braket{\phi}{\phi}
\end{equation}
with the following scalar product
\begin{equation}
\braket{\phi}{\psi}=\sum_{\rr} \Delta V\, \phi^*(\rr)\,\psi(\rr).
\end{equation}
From this splitting of the field operator it follows that
\begin{eqnarray}
\Psih(\rr)\ket{N:\phi}&=&\sqrt{N}\phi(\rr)\ket{N-1:\phi} \\
\Psihd(\rr)\ket{N:\phi}&=&\sqrt{N+1}\frac{\phi^*(\rr)}{\norme{\phi}^2}
\ket{N+1:\phi}+\Psihd_\perp(\rr)\ket{N:\phi}.
\end{eqnarray}
Inserting these expressions into the Schr\"odinger equation 
for the Hartree-Fock state  and collecting the terms containing
the different powers of $\Psihd_\perp$, we finally obtain
\begin{multline}
\Hamilt\ket{N:\phi}=\sqrt{N}\,\left(\sum_{\rr}\Delta
V\,\chi(\rr)\,\Psihd(\rr)\right)
\ket{N-1:\phi}\\
+\frac{\sqrt{N(N-1)}}{2}\,\left(\sum_{\rr}\Delta
V\,g_0\,\phi(\rr)\phi(\rr)\Psihd_\perp(\rr)\Psihd_\perp(\rr)\right)
\ket{N-2:\phi}
\eqname{SchrodPsi}
\end{multline}
with
\begin{equation}
\chi(\rr)=\left[h_0+\frac{g_0(N-1)}{\norme{\phi}^2}\valass{\phi(\rr)}^2-\frac{g_0(N-1)}{2\,\norme{\phi}^4}\sum_{\rr'}
\Delta V\,\valass{\phi(\rr')}^4\right]\phi(\rr).
\end{equation}

On the other hand, the variation of a Hartree-Fock state following a
variation of $\phi(\rr)$ can be calculated by replacing $\phi(\rr)$ by
$\phi(\rr)+d\phi(\rr)$ in \eq{Hartree} and expanding with the binomial
formula. Up to second order in the variation 
$d\phi(\rr)$, this leads to 
\begin{multline}
\ket{N:\phi+d\phi}=\ket{N:\phi}+
\sqrt{N}\left(\sum_{\rr}\Delta V\,d\phi(\rr)\,\Psihd(\rr)\right)\ket{N-1:\phi}
\\ +\frac{\sqrt{N(N-1)}}{2}\left(\sum_{\rr,\rr'}\Delta V^2
\,d\phi(\rr)\,d\phi(\rr')\,\Psihd(\rr)\,\Psihd(\rr')\right)\,\ket{N-2:\phi}+\ldots
\eqname{VarPhi}
\end{multline}

In the usual Hartree-Fock mean-field approximation, $\phi$ has a
deterministic evolution with $d\phi\propto dt$, so that the last term
on the right-hand side of \eq{VarPhi} is negligible in the limit
$dt\rightarrow 0$ and therefore the last term on the right-hand side of
\eq{SchrodPsi} cannot be accounted for. The best that one can do in
the mean-field approximation is to set
\begin{equation}
\left.d\phi(\rr)\right|_{\rm MF}=\frac{dt}{i\hbar}\chi(\rr),
\end{equation}
thus recovering a nonlinear Schr\"odinger equation for $\phi$ of the
same form as the usual Gross-Pitaevskii equation~\cite{RevueBEC}.

\begin{figure}[htbp]
\centerline{\includegraphics[width=10cm,clip=]{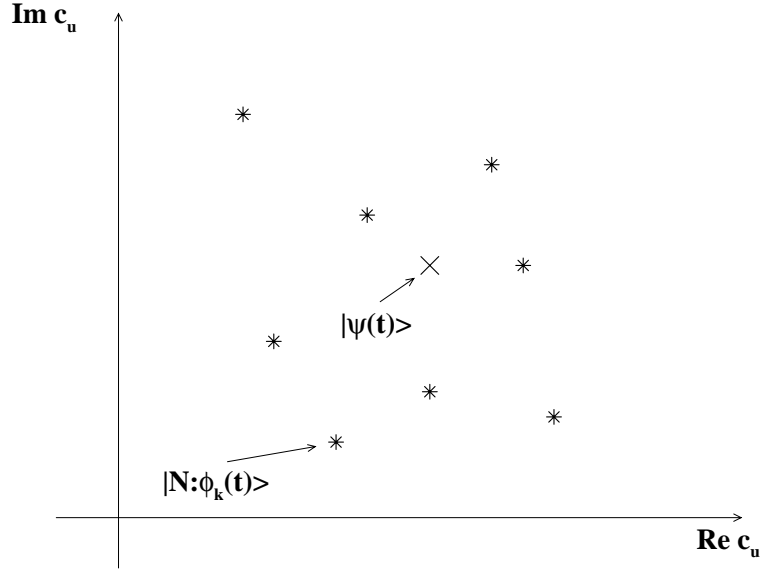}}
\caption{Schematic view of the Monte Carlo sampling of the exact
many-body state vector $\ket{\psi}$  at time $t$ (cross) by a collection of random
Hartree-Fock states $\ket{N:\phi_k}$ (stars). In the limit of an
infinite number of realizations, the center of mass of the
distribution of the stars exactly coincides with the cross.
In the $N$-body Hilbert space, $u$ is an arbitrary unit vector and
$c_u$ is the component along $u$ of $\ket{\psi}$ and of the $\ket{N:\phi_k}$.
\label{fig:figure}}
\end{figure}

The main idea to recover the exact evolution is to include in $d\phi$ a
term $dB$ proportional to $\sqrt{dt}$:
\begin{equation}
d\phi(\rr)=\frac{dt}{i\hbar}\chi(\rr)+dB(\rr)
\end{equation}
so that the quadratic term in $d\phi$ in \eq{VarPhi} is no longer negligible. 
The price to pay for this is the apparition of an extra term
proportional to $\sqrt{dt}$ in \eq{VarPhi} which has no counterpart
in \eq{SchrodPsi}. 
The trick is to make $dB$ stochastic with a vanishing
mean value. Then the average variation of the Hartree-Fock ansatz
exactly matches the one required by the many-body Schr\"odinger equation:
\begin{equation}
\mean{d\ket{N:\phi}}=\frac{dt}{i\hbar}\,\Hamilt\,\ket{N:\phi}
\end{equation}
provided that the noise term has the following correlation function
\begin{equation}
\mean{dB(\rr)dB(\rr')}=\frac{dt}{i\hbar}\proj_{\rr}\proj_{\rr'}
\left[\frac{g_0}{\Delta V}\delta_{\rr,\rr'}\phi(\rr)\phi(\rr')\right]
\eqname{CorrF}
\end{equation}
where the projector $\proj_{\rr}={\mathbf 1}-
\ketbra{\phi}{\phi}/\norme{\phi}^2$ projects orthogonally to $\phi$.
The zero mean property of the noise makes the terms proportional to
$\sqrt{dt}$ vanish in the averaging over noise. 
The terms of order three or more in $d\phi$ in the expansion
\eq{VarPhi} are negligible as compared to $dt$ in the limit
$dt\rightarrow 0$.
In mathematical terms, this procedure corresponds to assuming that the
evolution of $\phi$ is governed by an Ito stochastic differential
equation~\cite{StochMeth}.
Fig.\ref{fig:figure} gives a geometrical interpretation of the Monte
Carlo sampling of the many-body state vector $\ket{\psi}$ by
Hartree-Fock states.
The exact many-body state vector at any time $t>0$ $\ket{\psi(t)}$ is
equal to the statistical average of Hartree-Fock ansatz:
\begin{equation}
\ket{\psi(t)}=\big\langle\ket{N:\phi(t)}\big\rangle
\eqname{FockIdentity}
\end{equation}
provided that this is the case at the initial time $t=0$.


In practice, a noise with the correlation function \eq{CorrF} can be obtained as
\begin{equation}
dB(\rr)=\sqrt{\frac{dt\, g_0}{i\hbar V}}
\proj_\rr\left[\phi(\rr)\sum_{\kk>0}(e^{i(\kk\cdot\rr+\theta(\kk))}+\textrm{c.c.})\right];
\eqname{noise1}
\end{equation}
this expression was already discussed in~\cite{GPstoch} and is easily
generalized to finite range interaction potentials. 
The index $\kk$ is restricted to a half space and to the first
Brillouin zone. The $\theta(\kk)$'s are random angles uniformly 
distributed in $[0,2\pi]$.
An alternative choice can be
\begin{equation}
dB(\rr)=\sqrt{\frac{dt\, g_0}{i\hbar\,\Delta V}}
\proj_\rr\left[\phi(\rr)\,d\xi(\rr)\right]
\eqname{noise2}
\end{equation}
with the $d\xi(\rr)$ independent zero-mean random variables with 
$\mean{d\xi(\rr)\,d\xi(\rr')}=\delta_{\rr,\rr'}\,dt$.
In both cases
\begin{equation}
\mean{\norme{dB}^2}\leq \frac{g_0\,dt}{\hbar\,\Delta V}\,\norme{\phi}^2.
\end{equation}
We can then calculate the evolution of the squared norm
$\norme{\phi}^2$ and check that it remains finite at all times:
\begin{equation}
d\norme{\phi}^2=\mean{\norme{dB}^2}\leq \frac{g_0\,dt}{\hbar\,\Delta
V}\,\norme{\phi}^2,
\end{equation}
the absence of terms linear in $dB$ being due to the orthogonality
of $dB$ with respect to $\phi$.
This leads to
\begin{equation}
\norme{\phi}^2\!(t)\leq
\norme{\phi}^2\!(0)\;e^{\frac{g_0\,t}{\hbar\,\Delta V}},
\eqname{EvNorme}
\end{equation}
so that $\phi$ remains finite at all times.
The present stochastic wave function reformulation of the bosonic many-body
problem is exactly the same as the so-called {\em simple} scheme with
Fock states introduced in~\cite{GPstoch} from a different and more
general point of view.
Among the schemes based on a Hartree-Fock ansatz for the many-body density
matrix of the form 
\begin{equation}
\sigma(t)=\ketbra{N:\phi_1}{N:\phi_2}
\end{equation}
with $\phi_{1,2}$ evolving according to stochastic
differential equations, the {\em simple} Fock scheme was shown to be the one
which minimizes the growth rate of the statistical variance of $\sigma$
around the exact many-body density matrix $\rho(t)$:
\begin{equation}
\textrm{Var}[\sigma]\equiv\big\langle{\norme{\rho(t)-\sigma(t)}^2}
\big\rangle=
\big\langle\trace{\sigma^\dagger(t)\sigma(t)}\big\rangle
-\trace{\rho(t)^2},
\end{equation}
where $\big\langle\ldots\big\rangle$ represents the average over all
stochastic realizations.
The last term in the above equation, 
the so-called purity of $\rho$, is constant for
Hamiltonian evolutions.
If the stochastic wave functions $\phi_1$ and $\phi_2$ are statistically
independent at the initial time $t=0$, they remain so
at any later time since their evolutions are independent. The mean
squared norm of $\sigma$ is then 
\begin{equation}
\big\langle\trace{\sigma^\dagger(t)\sigma(t)}\big\rangle =
\langle\Delta\rangle^2(t)
\end{equation}
where 
\begin{equation}
\Delta(t)\equiv \langle N:\phi(t)|N:\phi(t)\rangle=\norme{\phi}^{2N}.
\end{equation}
A geometrical interpretation of $\Delta$ can be put forward in the
spirit of fig.\ref{fig:figure}: the mean value of $\Delta$ is related to the 
mean value of the squared distance in Hilbert space between the Monte
Carlo ansatz and the exact many-body state vector
\begin{equation}
\expect{\Delta}-\norme{\ket{\psi}}_H^2=\expect{\norme{\ket{N:\phi}-\ket{\psi}}_H^2}.
\end{equation}
For the present Fock scheme, it follows from \eq{EvNorme} that
\begin{equation}
\langle\Delta\rangle(t)\leq \langle\Delta\rangle(0)
\,e^{{ N g_0 t}/{\hbar \, \Delta V}};
\eqname{Delta}
\end{equation}
this upper bound was already discussed in~\cite{GPstoch}. 
In the model of this paper, the number of modes of the bosonic
field is finite, so that the inequality \eq{Delta} guarantees that
the simulation can predict the expectation value of any observable
with a finite statistical dispersion.
The Monte Carlo statistical variance $\Delta O^2$
on the quantum expectation value $\textrm{Tr}[\rho O]$ of 
any observable ${O}$ is indeed limited from above by
\begin{equation}
\Delta O^2\equiv\expect{|\textrm{Tr}[\sigma {O}]\,|^2}-\textrm{Tr}[\rho O]^2\leq
\textrm{Tr}[{O}^2]\,\expect{\textrm{Tr}[\sigma^\dagger \sigma]}-\textrm{Tr}[\rho O]^2
\end{equation}
where the traces are taken in the $N$-particle subspace and the trace
$\textrm{Tr}[{O}^2]$ is a finite quantity for a finite number of modes.

This formalism can be extended to the imaginary-time
evolution~\cite{GPstochT}, which allows one to sample in an exact way
the thermal
equilibrium state at a given temperature $T$:
\begin{equation}
\rho=e^{-\Hamilt/k_B T}.
\end{equation}
This requires solving the imaginary-time evolution equation
\begin{equation}
\frac{d}{d\tau}\rho=-\frac{1}{2}\{\Hamilt,\rho\}
\end{equation}
for a ``time'' $\tau$ going from $0$ to $\beta=1/k_B T$.
The use of the projector onto the $N$ particle subspace ${\mathcal
P}_N$ as the initial condition allows one to calculate the physical
quantities in the canonical ensemble, i.e. for a fixed number $N$ of
particles. As we shall prove in full detail in the next section, the
projector operator ${\mathcal P}_N$ has a simple expression in terms
of Hartree-Fock states. 

The correct imaginary-time evolution of the Hartree-Fock state
$\ket{N:\phi}$ is recovered if the stochastic wave function $\phi(\rr)$
evolves according to 
\begin{equation}
d\phi(\rr)=-\frac{d\tau}{2}\left[h_0+\frac{g_0(N-1)}{\norme{\phi}^2}\valass{\phi(\rr)}^2-\frac{g_0(N-1)}{2\,\norme{\phi}^4}\sum_{\rr'}
\Delta V\,\valass{\phi(\rr')}^4\right]\phi(\rr)+dB(\rr)
\eqname{stoch}
\end{equation}
with a noise term $dB(\rr)$ equal to
\begin{equation}
dB(\rr)=i\sqrt{\frac{d\tau\, g_0}{2V}}
\proj_\rr\left[\phi(\rr)\sum_{\kk>0}(e^{i(\kk\cdot\rr+\theta(\kk))}+\textrm{c.c.})\right].
\eqname{dBImT}\end{equation}
We have recently applied this method to study the probability distribution
of the number of condensate particles in a 1D trapped
Bose gas for temperatures both above and below the transition 
temperature \cite{StatN0}.

\section{Proof of (over)-completeness of the Hartree-Fock states}
\label{sec:YvanTh}

In the previous section we have shown how the $N$-boson time-dependent
problem can be
exactly reformulated in terms of the stochastic evolution of single particle
wave functions; in particular, the thermal
equilibrium state in the canonical ensemble can be obtained by means
of an imaginary-time evolution with the projector ${\mathcal P}_N$ as
the initial state.
Here, we shall prove that the projector ${\mathcal P}_N$ has a
simple expression in terms of Hartree-Fock states and that any state vector
can be expanded onto a set of Hartree-Fock states. This ensures that
the identity \eq{FockIdentity} can be satisfied at time $t=0$ for any
initial many-body state vector $\ket{\psi(0)}$.

It is a well-known fact of quantum optics~\cite{QO} that the Glauber coherent
states form an overcomplete basis of the Hilbert space of an harmonic
oscillator. In the general multi-mode case, this property can be
expressed as
\begin{equation}
{\mathbf 1}=\frac{1}{\pi^M}\int\!{\mathcal D}\phi\,\ketbra{\textrm{Glaub}:\phi}{\textrm{Glaub}:\phi}
\eqname{1coh}
\end{equation}
where the Glauber coherent state is defined as usual as
\begin{equation}
\ket{\textrm{Glaub}:\phi}=e^{-\norme{\phi}^2/2}\,e^{\sum_\rr\Delta
V\,\phi(\rr)\Psihd(\rr)}\ket{0}.
\end{equation}
and ${\mathbf 1}$ is the identity on the bosonic subspace of totally
symmetric wave functions.
The integration in \eq{1coh} is performed over the $2M$ dimensional
space of complex-valued functions defined on the $M$-point
lattice:
\begin{equation}
{\mathcal D}\phi=\prod_\rr d\textrm{Re}\phi(\rr)\,d\textrm{Im}\phi(\rr).
\end{equation}
Using the well-known formula
\begin{equation}
{\mathcal P}_N\ket{\textrm{Glaub}:\phi}=\frac{1}{\sqrt{N!}}e^{-\norme{\phi}^2/2}\ket{N:\phi}
\end{equation}
to project onto the $N$ particle subspace, an explicit expression for
${\mathcal P}_N$ in terms of Hartree-Fock states can be obtained
\begin{equation}
{\mathcal P}_N={\mathcal P}_N {\mathbf 1} {\mathcal
P}_N=\frac{1}{\pi^M}\int\!{\mathcal D}\phi\,{\mathcal P}_N
\ketbra{\textrm{Glaub}:\phi}{\textrm{Glaub}:\phi}{\mathcal
P}_N=C\,\int_1\!{\mathcal D}{\hat \phi}\,\ketbra{N:{\hat
\phi}}{N:{\hat \phi}}.
\eqname{PN}
\end{equation}
The constant $C$ is equal to
\begin{equation}
C\equiv\frac{1}{N!\,\pi^M} \int\! d\norme{\phi}\,\norme{\phi}^{2N+2M-1}
e^{-\norme{\phi}^2}=\frac{(M+N-1)!}{2\,\pi^M\,N!}.
\end{equation}
The normalized wave function ${\hat \phi}(\rr)$ is defined as ${\hat \phi}(\rr)=\phi(\rr)/\norme{\phi}$ and the integration over ${\hat
\phi}$ is performed on the unit sphere $\|\hat \phi\|^2=1$ 
according to the measure ${\mathcal D}{\hat \phi}$ defined as
\begin{equation}
{\mathcal D}\phi=\norme{\phi}^{2M-1}\,d\norme{\phi}\,{\mathcal D}{\hat \phi}.
\end{equation}
As discussed in~\cite{GPstochT}, the expression \eq{PN} is the
starting point for an exact numerical calculation of the thermodynamical
equilibrium properties of the interacting Bose gas.

Completeness of the set of Hartree-Fock states is an immediate
consequence of the result \eq{PN}: any state vector $\ket{\psi}$ belonging to
the $N$ particle subspace can in fact be expanded over Hartree-Fock states
according to
\begin{equation}
\ket{\psi}={\mathcal P}_N\ket{\psi}=C\,\int_1\!{\mathcal
D}{\hat \phi}\,\braket{N:{\hat \phi}}{\psi}\;\ket{N:{\hat \phi}}.
\end{equation}
As required for practical implementation of the Fock scheme,
 the weight factor can be made real
and positive for all the
${\hat \phi}$ just by reabsorbing the phase of $\braket{N:{\hat
\phi}}{\psi}$ into that of the wave function ${\hat \phi}$:
\begin{equation}
\ket{\psi}=C\,\int_1\!{\mathcal
D}{\hat \phi}\,\valass{\braket{N:{\hat \phi}}{\psi}}\;
\ket{N:{r({\hat \phi})}\,{{\hat \phi}}}
\end{equation}
with the phase factor $r({\hat \phi})$ defined according to
\begin{equation}
r({\hat \phi})^N=\frac{\braket{N:{\hat
\phi}}{\psi}}{\valass{\braket{N:{\hat \phi}}{\psi}}}.
\end{equation}
Since the scalar product of two Hartree-Fock states is equal to
\begin{equation}
\braket{N:\phi_2}{N:\phi_1}=\braket{\phi_2}{\phi_1}^N,
\end{equation}
this family of states is not an orthonormal basis, but rather an
overcomplete set.

\section{Stochastic schemes with coherent states: an infinite
statistical uncertainty}

\subsection{The Bargmann scheme}

The derivation of the Fock scheme can be reproduced for the case of 
unnormalized coherent states, the so-called Bargmann states:
\begin{equation}
|\textrm{Barg}:\phi\rangle=\Pi_0\,e^{\sum_\rr\!\Delta V\,\phi(\rr)\Psihd(\rr)}\ket{0},
\eqname{BargStates}
\end{equation}
where $\phi$ is a stochastic dynamical variable and $\Pi_0$ is a
time-independent amplitude.
Thanks to \eq{1coh}, one can show that any initial state vector
$\ket{\psi(0)}$ can be obtained as a statistical average of Bargmann
states of the form \eq{BargStates}; the initial value of both $\phi$
and $\Pi_0$ can then vary randomly from one Monte Carlo realization to
the other.

The exact time-evolution given by the many-body Hamiltonian \eq{Hamilt}
can be shown to be exactly recovered if the field amplitude
$\phi(\rr)$ evolves according to
\begin{equation}
d\phi(\rr)=\frac{dt}{i\hbar}h_0\phi(\rr)+dB(\rr)
\eqname{B1}
\end{equation}
with a noise correlation function equal to
\begin{equation}
\mean{dB(\rr)\,dB(\rr')}=\frac{dt}{i\hbar}\,\frac{g_0}{\Delta V}\,\delta_{\rr,\rr'}\,\phi(\rr)\phi(\rr').
\eqname{B2}
\end{equation}
In practice, we take the following implementation of the noise
which minimizes the squared norm of $dB$ \cite{GPstoch}:
\begin{equation}
dB(\rr)=\sqrt{\frac{dt\, g_0}{i\hbar V}}
\phi(\rr)\sum_{\kk>0}(e^{i(\kk\cdot\rr+\theta(\kk))}+\textrm{c.c.}),
\eqname{noise3}
\end{equation}
where $\kk$ is restricted to a half space and to the first Brillouin zone.
Within the framework of~\cite{GPstoch}, this Bargmann scheme
recovers one of
the possible solutions of the consistency equations in the case of a
coherent state ansatz, namely the one with a constant amplitude 
$\Pi(t)$ at all times.

Here we shall prove that, contrary to the scheme with Fock states, the
scheme (\ref{eq:B1}-\ref{eq:B2}) has an infinite statistical
uncertainty for
any time $t>0$, in a sense to be precised below. 
As in the section with Fock states we introduce
\begin{equation}
\Delta(t)\equiv \langle\textrm{Barg}:\phi(t)|\textrm{Barg}:\phi(t)\rangle
=\valass{\Pi_0}^2\,e^{\norme{\phi(t)}^2}.
\eqname{Dt}
\end{equation}
As initial state at $t=0$ we take for simplicity
a pure Bargmann state $|\textrm{Barg}:\phi(0)\rangle$ with in 
average $\norme{\phi(0)}^2=N$ particles. $\Pi_0$ is taken as
$\exp(-N/2)$, so that the initial state is normalized to unity.
We now show that the mean value of $\Delta(t)$ is infinite for any $t>0$.

From \eq{B1} and \eq{B2}, it follows that the evolution of
$X(t)=\norme{\phi(t)}^2$ is described by
\begin{equation}
dX=\braket{dB}{dB}+\braket{dB}{\phi}+\braket{\phi}{dB}=\frac{g_0}{\hbar\,\Delta V}\,X\,dt+d\eta
\eqname{dX2}
\end{equation}
where  
\begin{equation}
d\eta=\braket{dB}{\phi}+\braket{\phi}{dB}
\end{equation}
is a zero-mean noise term and its variance is equal to
\begin{equation}
\mean{d\eta^2}=\mean{(\braket{dB}{\phi}+\braket{\phi}{dB})^2}=\mean{\braket{\phi}{dB}^2}+\mean{\braket{dB}{\phi}^2}+2\mean{\braket{dB}{\phi}\braket{\phi}{dB}}.
\eqname{eta}
\end{equation}
Inserting in this expression the noise correlation function \eq{B2},
the first two terms on the right-hand side of \eq{eta} cancel.
For the remaining term, a lower bound can be obtained by using the
explicit expression \eq{noise3} for the noise and applying the Schwartz inequality
\footnote{For a fixed value of the norm of the field $\norme{\phi}$, the sum
$\sum_\rr |\phi(\rr)|^4$ is minimum if the density $|\phi(\rr)|^2$ is
a constant, which implies $\sum_\rr \Delta V\, |\phi(\rr)|^4\geq
\norme{\phi}^4/M\,\Delta V$.}:
\begin{equation}
\mean{d\eta^2}=2\mean{\braket{dB}{\phi}\braket{\phi}{dB}}=\frac{2g_0\,dt}{\hbar}\sum_\rr\Delta
V\,\valass{\phi(\rr)}^4\geq
\frac{2}{M} \frac{g_0}{\hbar\,\Delta V}\,X^2\,dt;
\end{equation}
as in the previous sections, $M$ is the number of lattice points.

From \eq{dX2}, a lower bound on the increase of the moments of
$\big\langle\, X^n\big\rangle(t)$ can be obtained:
\begin{multline}
d\big\langle X^n\big\rangle=n\,\big\langle X^{n-1}\,dX\big\rangle
+\frac{n(n-1)}{2}\,\big\langle
X^{n-2}\,dX^2\big\rangle=\frac{g_0\,dt}{\Delta
V\,\hbar}\,n\,\big\langle X^n\big\rangle+
\frac{n(n-1)}{2}\big\langle X^{n-2}
\,d\eta^2 \big\rangle 
\geq
\\ \geq \frac{g_0\,dt}{\Delta
V\,\hbar}\left[n\,\big\langle
X^n\big\rangle+\frac{n(n-1)}{M}\,\big\langle X^n\big\rangle\,\right]
\end{multline}
and therefore on the moments themselves:
\begin{equation}
\big\langle X^n\big\rangle(t)\geq \big\langle X^n\big\rangle(0)\,e^{\gamma(n)\tau}=N^n\,e^{\gamma(n)\tau}.
\eqname{XN}
\end{equation}
Here we have set $\tau=g_0 t/\Delta V \hbar$ and
\begin{equation}
\gamma(n)=n+n(n-1)/M.
\end{equation}
This leads to the lower bound for the mean value
of $\Delta(t)$:
\begin{equation}
\langle\Delta\rangle(t)=\valass{\Pi_0}^2\,\big\langle\exp{X}\big\rangle
\geq \valass{\Pi_0}^2\,\sum_{n=0}^\infty\frac{1}{n!}\,N^n
e^{\gamma(n)\tau}.
\eqname{DeltaSum}
\end{equation}
As the ratio of two successive elements of the series
tends to infinity in the limit $n\rightarrow\infty$ for any $t>0$,
this results in an infinite value of $\langle\Delta\rangle(t)$
and therefore of the variance of the statistical error on the many-body density 
operator.

The same result could have been obtained by explicitly calculating the
probability distribution for $X$ at a given time $t$: as the evolution
of the random variable $Y=\ln(X)$ is a simple diffusive motion
in presence of a constant drift term, its
probability distribution $P(Y,t)$ has a Gaussian shape at all times
$t$. 
By going back to the original variable $X$ by means of
$P(X,t)=\frac{dY}{dX}P(Y,t)$, the distribution of $X$ is found to be equal to
\begin{equation}
P(X,t)=\frac{1}{X}\,\left(\frac{M}{4\pi\tau}\right)^{1/2}
\,e^{-\frac{M}{4\tau}\left[\ln(X/N)-(1-\frac{1}{M})\tau\right]^2}.
\eqname{ProbX}
\end{equation}
It follows that the average of the exponential
$\exp(X)$ does not converge at $X=+\infty$
and therefore that the average of \eq{Dt} is indeed infinite.

One might wonder if the same conclusion holds for other indicators of
the magnitude of the statistical error. For example, one could
consider the mean value of the distance $\delta$ between the Monte Carlo ansatz 
$|\textrm{Barg}:\phi\rangle$ and the exact many-body state vector
$\ket{\psi}$:
\begin{equation}
\delta=\norme{|\textrm{Barg}:\phi\rangle-\ket{\psi}}_H
\end{equation}
where $\norme{\ldots}_H$ is the usual norm in the many-body Hilbert
space.
As
\begin{equation}
\norme{|\textrm{Barg}:\phi\rangle}_H\leq\norme{|\textrm{Barg}:\phi\rangle-\ket{\psi}}_H+
\norme{\ket{\psi}}_H
\end{equation}
the mean distance $\expect{\delta}$ is bounded from below by
\begin{equation}
\expect{\delta}\geq \expect{\Delta^{1/2}}-\norme{\ket{\psi}}_H.
\end{equation}
By a  straightforward generalization of \eq{DeltaSum}, we find that
the mean value of $\Delta^{1/2}$ is infinite for any $t>0$, so that
the mean value of $\delta$ is infinite.
From the explicit form of the probability distribution \eq{ProbX}
one can also check that the same result holds for any moment
$\expect{\delta^\alpha}$ of the distance $\delta$, with an arbitrary
real exponent $\alpha>0$.

These results nicely demonstrate the difference between the
convergence of the stochastic process for the field amplitude 
$\phi$ and for the many-body ansatz.
Since the deterministic term \eq{B1} is norm-conserving and the noise term
\eq{B2} does not grow faster than $\norme{\phi}$, one can apply a
regularity theorem which guarantees that no finite time divergences 
can occur in the random dynamics
of the field amplitude $\phi$ and that all the moments of the stochastic wave function
are finite (cf. \cite[chap. 4.3]{StochMeth} and \cite[chap.6-7]{StochAnal}).
However, the statistical uncertainty is still divergent because the
Bargmann ansatz involves an exponential function of
the field amplitude $\phi(\rr)$ and not only a polynomial one as was
the case with Fock states~\cite{GPstoch}.

\subsection{The {\em simple} scheme with coherent states}

In the previous subsection we showed that the statistical uncertainty for a
simulation performed according to the Bargmann scheme is
infinite. 
In~\cite{GPstoch} we have identified  among all the schemes based on coherent
states the one (the so-called {\em simple} scheme) which minimizes the growth
rate of the variance of the statistical error on the many-body density
operator. Here we shall show that for any finite time the statistical
uncertainty is infinite  also for the {\em simple} scheme,
similarly to the case of the Bargmann scheme.

The scheme denoted in~\cite{GPstoch} as the {\em simple} scheme with
coherent states corresponds to a slightly more general ansatz
than \eq{BargStates}:
\begin{equation}
\ket{\textrm{coh}:(\phi,\Pi)}=\Pi\,e^{\sum_\rr\!\Delta V\,\phi(\rr)\Psihd(\rr)}\ket{0}
\eqname{AnsSC}
\end{equation}
where the amplitude $\Pi(t)$ is now a stochastically evolving dynamical variable.
The field amplitude $\phi$ now solves the Ito differential stochastic
equation
\begin{equation}
d\phi(\rr)=\frac{dt}{i\hbar}\left[h_0+g_0\valass{\phi(\rr)}^2\right]\phi(\rr)
+dB(\rr)
\eqname{SC1}
\end{equation}
with exactly the same noise term $dB$ as in the Bargmann scheme,
see \eq{noise3}.
The evolution of the amplitude $\Pi$ is given by
\begin{equation}
d\Pi=-\Pi\,\sum_\rr\Delta V\,\phi^*(\rr)dB(\rr)=-\Pi\,\braket{\phi}{dB}.
\eqname{SC2}
\end{equation}
The deterministic term in $d\phi$ includes the usual mean field
contribution. As it remains norm conserving,
the results previously proved for the distribution and the moments of
$X(t)=||\phi||^2$ for the Bargmann scheme still hold for the present 
scheme. 
In particular, the stochastic differential equation \eq{SC1} has a
well defined solution according to the regularity theorem of
\cite{StochMeth,StochAnal}.
Setting $\Pi=\exp S$, equation \eq{SC2} can be turned into 
\begin{equation}
dS=-\braket{\phi}{dB}-\frac{1}{2}\braket{\phi}{dB}^2.
\end{equation}
Using the explicit expression for the correlation function of $dB$,
this expression simplifies to
\begin{equation}
dS=-\braket{\phi}{dB}-\frac{g_0\,dt}{2i\hbar}\sum_\rr\Delta V\, \valass{\phi(\rr)}^4.
\end{equation}
This equation for $S$ can be integrated by quadrature proving the
existence of a regular solution to \eq{SC2}.

As in the previous section we quantify the statistical spread in
the simulation with
\begin{equation}
\Delta(t)\equiv \langle\mbox{coh}:(\phi,\Pi)|\mbox{coh}:(\phi,\Pi)\rangle.
\end{equation}
As the initial state at $t=0$, we take for simplicity a pure
coherent state $\ket{\textrm{coh}:(\phi(0),\Pi(0))}$ with on average
$\norme{\phi(0)}^2=N$ particles and an amplitude $\Pi(0)=\exp(-N/2)$, so
that the state vector is normalized to unity.

The time evolution of $\Delta(t)$  is given by~\cite{GPstoch}:
\begin{equation}
d\Delta=||dB||^2 \Delta =\frac{g_0}{\hbar\,\Delta V}\,X\,\Delta\,dt.
\eqname{dDelta}
\end{equation}
Because of the nonlinear term proportional to $X(t)\Delta(t)$ in
\eq{dDelta}, the coupled system of \eq{dDelta} and \eq{dX2} does not
fulfill all the hypothesis of the regularity
theorem~\cite[chap. 4.3]{StochMeth} and therefore we can expect
singular behaviors.

In terms of the rescaled time $\tau=g_0 t/\Delta V\,\hbar$, the
stochastic equation of motion for $X(\tau)$ and $\Delta(\tau)$ can be
rewritten as
\begin{eqnarray}
dX&=&X\,d\tau+d\eta \\
d\Delta&=&X\,\Delta\,d\tau, \eqname{DeltaTau} 
\end{eqnarray}
where the zero-mean noise $d\eta$ satisfies, as in the previous section:
\begin{equation}
\mean{d\eta^2}\geq\frac{2}{M}X^2\,d\tau.
\end{equation}
It follows from \eq{DeltaTau} that
\begin{equation}
\Delta(t)=\exp\left({\int_0^\tau\!d\tau'\,X(\tau')}\right)
\end{equation}
and therefore its mean value can be written as the series
\begin{equation}
\langle\Delta\rangle(\tau)=1+\sum_{n=1}^\infty D_n(\tau)
\eqname{series}
\end{equation}
whose coefficients $D_n(\tau)$ are given by integrals of the form
\begin{equation}
D_n(\tau)=\int_{\tau>\tau_n>\tau_{n-1}>\ldots>\tau_2>\tau_1>0}
\hspace{-3.3cm}d\tau_n\,d\tau_{n-1}\ldots
d\tau_2\,d\tau_1\,\big\langle X(\tau_n)X(\tau_{n-1})\ldots
X(\tau_2)X(\tau_1)\big\rangle.
\eqname{T-ord}
\end{equation}
Using the regression theorem of stochastic analysis~\cite{StochMeth} and the fact
 that all moments \eq{XN} of $X$ are increasing functions of $\tau$, 
it is easy to see that
\begin{equation}
\big\langle X^k(\tau)X(\tau_{n-k})X(\tau_{n-k-1})\ldots
X(\tau_{2})X(\tau_{1})\big\rangle
\end{equation}
is itself an increasing function of $\tau$ for any $1\leq
k\leq n$ and for $\tau\geq\tau_{n-k}\geq\tau_{n-k-1}\geq\ldots\geq\tau_1\geq 0$.
This implies that
\begin{multline}
\big\langle{X(\tau_n)\,X(\tau_{n-1})\,X(\tau_{n-2})\ldots X(\tau_1)}\big\rangle\geq\\
\geq \big\langle{X(\tau_{n-1})^2\,X(\tau_{n-2})\ldots
X(\tau_1)}\big\rangle \geq\big\langle{X(\tau_{n-2})^3\ldots
X(\tau_1)}\big\rangle\geq\ldots\geq\big\langle{X(\tau_1)^n}\big\rangle\geq N^n\,e^{\gamma(n)\tau_1}.
\end{multline}
For $\tau>0$, we then have
\begin{equation}
D_n(\tau)\geq \frac{N^n\,e^{\gamma(n)\tau}}{(n-1)!\,\gamma(n)^n}\,\int_0^{\gamma(n)\tau}\!e^{-w}\,w^{n-1}\,dw.
\end{equation}
The integral in $w$ can be explicitly performed and for values of $n$
sufficiently large that $\gamma(n)\tau\geq 1$, it turns out to be bounded from 
below as follows:
\begin{equation}
\int_0^{\gamma(n)\tau}\!e^{-w}\,w^{n-1}\,dw=(n-1)!\left[1-e^{-\gamma(n)\tau}\,\sum_{j=0}^{n-1}\frac{(\gamma(n)\tau)^j}{j!}\right]
\geq(n-1)!\,\left[1-n(\gamma(n)\tau)^{n-1}\,e^{-\gamma(n)\tau}\right]
\end{equation}
which means that
\begin{equation}
D_n\geq \frac{N^n\,e^{\gamma(n)\tau}}{\gamma(n)^n}\left[1-n\,\left(\gamma(n)\tau\right)^{n-1}\,e^{-\gamma(n)\tau}\right]\equiv E_n.
\end{equation}
Since
\begin{equation}
\lim_{n\rightarrow \infty} \frac{E_{n+1}}{E_n}\equiv+\infty
\end{equation}
the infinite sum of the $E_n$ is infinite as well as the sum of the
$D_n$ which appears in \eq{series}.
This proves that the variance of the statistical error on the
many-body density
matrix is indeed infinite also in the case of the simple scheme with coherent states. 

As for the Bargmann scheme, this result can be generalized to other
indicators of the statistical error, e.g. the distance $\delta$
between the Monte Carlo ansatz and the exact state vector
\begin{equation}
\delta=\norme{|\textrm{coh}:(\phi,\Pi)\rangle-\ket{\psi}}_H
\end{equation}
has an infinite mean for $t>0$.

\section{Conclusion}
We have shown that the {\em simple} Fock scheme can be rederived
in a very elementary way using the Schr\"odinger equation for
the state vector rather than the Liouville equation for the
many-body density operator. Such a derivation is inspired
from the derivation for fermions
used in \cite{Chomaz}. We have also given
a proof of completeness of the Hartree-Fock state ansatz more detailed
than in \cite{GPstoch}.

For the two schemes with coherent states considered in this paper,
we have shown that the absolute value of the statistical error on the many-body
state vector has an infinite mean for any finite evolution time.
This negative result has been obtained by analytical means for a continuous
time evolution. Of course,
any actual numerical calculation is performed with a
finite time step $dt$ so that any given time $t$ is reached
with a finite number of steps.
As the noise \eq{noise3}  is bounded from
above, this imposes an upper limit to the maximum value of the
statistical error which can
be achieved by any realization and therefore its mean value
remains finite in the numerics.
This is why this problem of the {\em simple} coherent
scheme was not discovered in the numerical simulations
of \cite{GPstoch}.

\begin{acknowledgments}
We acknowledge useful discussions with Crispin Gardiner, Karen
Kheruntsyan, Peter Drummond, Slava Yukalov, Reinhold Walser, Eugene
Zaremba, Carlos Lobo, Dimitri Gangardt, Alice Sinatra, Philippe Chomaz
and Olivier Juillet.
I.C. acknowledges a Marie Curie grant from the EU under contract number
HPMF-CT-2000-00901.
Laboratoire Kastler Brossel is a Unit\'e de
Recherche de l'\'Ecole Normale Sup\'erieure et de l'Universit\'e Paris
6, associ\'ee au CNRS.
\end{acknowledgments}








\end{document}